\documentclass[11pt,a4paper]{article}

\usepackage[utf8]{inputenc}
\usepackage{amsmath,amssymb,mathtools,bm}

\usepackage{booktabs}
\usepackage{graphicx}
\usepackage{xcolor}
\usepackage{subcaption}

\usepackage{geometry}
\usepackage{enumitem}
\graphicspath{{figures/}}

\usepackage[numbers,sort&compress]{natbib}
\usepackage[colorlinks=true,allcolors=blue!55!black]{hyperref}

\newcommand{\ii}{\mathrm{i}}
\newcommand{\EN}{E_{\mathcal N}}
\newcommand{\CM}{\boldsymbol\sigma}
\newcommand{\trans}{\mathsf T}
\newcommand{\tr}{\operatorname{tr}}
\newcommand{\diag}{\operatorname{diag}}
\newcommand{\atantwo}{\operatorname{atan2}}

\title{Exact Gaussian Entanglement Dynamics and Initial-State Control\\
in Coupled Parametric Oscillators}

\author{%
 Roberto Bernal-Jaquez\thanks{Corresponding author: \texttt{rbernal@cua.uam.mx}}\,$^{a}$,\quad
 Hector H. Hernandez-Hernandez$^{\,b,a}$ \\ 
   Alexander Schaum$^{\,c,d}$ \quad
   Guillermo Chac\'on-Acosta$^{\,a}$\\ [8pt]
 \small $^{a}$Departamento de Matem\'aticas Aplicadas y Sistemas,\\[-3pt]
 \small Universidad Aut\'onoma Metropolitana--Cuajimalpa, Ciudad de M\'exico, M\'exico\\[3pt]
 \small $^{b}$Facultad de Ingenier\'ia, Universidad Aut\'onoma de Chihuahua, Chihuahua, M\'exico\\ [3pt]
 \small $^{c}$Department for Process Analytics, University of Hohenheim, Stuttgart, Germany\\[3pt]
 \small $^{d}$Computational Science Hub, University of Hohenheim, Stuttgart, Germany}
\date{}

\begin{document}
\maketitle

\begin{abstract}
We study two bilinearly coupled harmonic oscillators driven by a common
time-dependent spring.  A fixed normal-mode transformation reduces the pair
to two parametric oscillators, each solved exactly by its Lewis-Riesenfeld
invariant and Ermakov-Pinney amplitude, including the ladder-operator
construction and the Lewis-Riesenfeld phase.  For invariant-vacuum inputs the
physical two-mode covariance matrix is obtained in closed form, and the
logarithmic negativity collapses to \(\EN=\tfrac12\operatorname{arcosh}(X/2)\),
where a single dimensionless invariant \(X\geq2\) measures the relative
phase-space deformation of the two normal modes.  We give a
normalization-explicit account of the Duan--Simon comparison: after local
symplectic optimization the Duan test and the positive partial transpose share
the entanglement threshold for this pure exchange-symmetric Gaussian family, but
the optimized Duan quantity is a witness, not an independent measure, with
\(\mathcal D_{\rm opt}/\mathcal D_{\rm sep}=e^{-\EN}\).  We carefully separate a
genuinely separable product preparation from the already-entangled ground state
of the coupled Hamiltonian, and use only the former for claims of entanglement
generation.  Under sinusoidal stiffness modulation a separable input shows
bounded correlations away from resonance and sustained growth near parametric
resonance, and we demonstrate numerically that the initial width and chirp act
as two independent, experimentally accessible controls of the entanglement at a
chosen target time.  An \(SU(1,1)\) formulation identifies \(2\EN\) with the
hyperbolic separation of the two normal-mode squeezing trajectories on the
\(SU(1,1)/U(1)\) disk and shows that entanglement is governed by both the
relative squeezing magnitude and the relative squeezing angle; we exhibit
regimes in which nearly equal magnitudes still produce strong entanglement
through the angle alone.  The construction extends to \(N\) oscillators sharing a
common modulation and a fixed coupling topology.
\end{abstract}


\section{Introduction}
\label{sec:intro}

Time-dependent harmonic systems are a rare meeting point of exact nonstationary
quantum dynamics, continuous-variable (CV) quantum information, and quantum
control.  They model the dynamical degrees of freedom of trapped ions,
mechanical modes of optomechanical resonators, and tunable superconducting
resonators~\cite{Leibfried2003,Aspelmeyer2014,Devoret2013}.  A time variation of
the quadratic confinement drives Bogoliubov mixing and squeezing, while a
coupling between modes converts a suitable relative squeezing into
bipartite entanglement.  The question we address is the following:
 for a fixed, experimentally natural coupling, how much
entanglement can a common frequency modulation create between two oscillators,
how is it distributed in time, and how can it be steered to a chosen value by
preparing the initial state alone?

The Lewis-Riesenfeld (LR) invariant~\cite{LewisRiesenfeld1969} supplies an
exact, nonperturbative and nonadiabatic basis for a single parametric
oscillator.  Its auxiliary amplitude obeys the Ermakov-Pinney
equation~\cite{Pinney1950}, and that one scalar function simultaneously fixes the
wavefunction width, its quadratic phase (chirp), and hence the entire Gaussian
covariance matrix.  Exact invariants and transformations for coupled
time-dependent oscillators have been constructed in several settings: for
arbitrary interactions through orthogonal-function invariants~\cite{Urzua2019},
for boundary-value inverse engineering~\cite{Tobalina2020}, and for three-mode
and quenched models~\cite{Park2019,Abidi2021}.  A recent study solves two
oscillators with a time-dependent coupling and reports purity and linear entropy
for excited states~\cite{Ghaba2026}.  Our
contribution is  complementary: we hold the bilinear coupling
constant, modulate the common stiffness, and ask for the exact Gaussian
entanglement and for its control through the initial width and chirp.  Our
comparison with the literature is about the control variable and the
entanglement diagnostic, not about whether the coupled time-dependent oscillator
can be solved.

Three conceptual points require care and are treated explicitly below.  First,
the instantaneous ground state of the coupled Hamiltonian is generically
entangled in the physical partition and cannot serve as a separable
initial condition for a generation claim; we introduce a genuinely separable
product preparation for that purpose and retain the coupled ground state only as
a benchmark (Sec.~\ref{sec:initial}).  Second, logarithmic negativity is a
quantitative entanglement parameter, whereas the Duan variance inequality is an
operational inseparability indicator; the two share a threshold for the pure
symmetric family studied here but are not the same object, and we make the
normalization relating them explicit (Sec.~\ref{sec:duan}).  Third,
\([I(t),H(t)]\neq0\) is a useful algebraic signature of the mismatch between the
invariant and energy bases, not by itself a dynamical cause: a static
Hamiltonian prepared with a squeezed invariant also has noncommuting \(I\) and
\(H\).  The physical sources of the dynamics are the parametric modulation, the
coupling, and the chosen initial state.

Our results connect to, and depart from, several recent lines of work.
\citet{Mirkhalaf2026} showed that in the static ground state of two
linearly coupled oscillators the frequency shift is dual to two-mode squeezing
and already witnesses entanglement at zero temperature; our closed form
\(\EN=\tfrac12\operatorname{arcosh}(X/2)\) carries this frequency--squeezing
duality into the fully driven regime, with \(X(t)\) acting as a time-dependent
squeezing-mismatch invariant between the normal modes.  \citet{Ghaba2026} treat
the complementary configuration, fixed frequencies with a time-dependent
coupling, diagnosed through purity and linear entropy, whereas our
constant-coupling, common-modulation setting yields a closed-form logarithmic
negativity governed by a single invariant.  \citet{GonzalezHenao2015} generate
and suppress entanglement in an open, bath-coupled parametric-oscillator
pair through the classical (Floquet) instability of the drive; our closed,
unitary system instead ties entanglement to the mismatch of the two normal-mode
Ermakov trajectories and exerts control through the initial Gaussian state.
\citet{Buchmann2018} establish symplectic controllability of a harmonic chain by
tuning the couplings in time; we keep the coupling fixed and control the
entanglement through the initial state.  Finally, our
initial-state control landscape is the dual of the boundary-value inverse
engineering of \citet{Tobalina2020}: there \(\Omega(t)\) is designed to meet
prescribed boundary conditions on the dynamical variables, whereas here
\(\Omega(t)\) is fixed and the initial (boundary) data are optimized.

The main results are: (i) an exact LR and ladder-operator solution for the two
normal modes, with closed-form covariance-matrix entries in terms of the Ermakov
amplitudes and their derivatives; (ii) a closed logarithmic negativity governed
by a single relative invariant \(X\), together with a corrected,
normalization-explicit comparison with the optimized Duan witness,
\(\mathcal D_{\rm opt}/\mathcal D_{\rm sep}=e^{-\EN}\); (iii) a control landscape
built on a genuinely separable input, showing that the initial width and
chirp are two independent handles on the target-time entanglement; (iv) an
\(SU(1,1)\) formulation in which \(2\EN\) is a hyperbolic distance and the
entanglement is set by both the relative squeezing magnitude and the relative
squeezing angle, including regimes in which nearly equal magnitudes still
entangle through the angle alone; and (v) a scalable extension to \(N\)
oscillators with common modulation and fixed coupling topology.  

\section{Model and exact Schr\"odinger solution}
\label{sec:model}

\subsection{Hamiltonian and fixed normal modes}

We set the mass of the oscillators to unity and keep \(\hbar\) explicit.  The Hamiltonian of the
two bilinearly coupled oscillators is
\begin{equation}
	H(t)=\frac12\sum_{j=1}^{2}\left[p_j^2+\omega^2(t)\,x_j^2\right]
	+\lambda\,x_1x_2,
	\qquad [x_j,p_k]=\ii\hbar\,\delta_{jk},
	\label{eq:H}
\end{equation}
with a common time-dependent stiffness \(\omega^2(t)\) and a constant bilinear
coupling \(\lambda\).  The quadratic potential is confining as long as
\(\omega^2(t)>|\lambda|\).  The time-independent canonical 
transformation
\begin{equation}
	Q_\pm=\frac{x_1\pm x_2}{\sqrt2},\qquad
	P_\pm=\frac{p_1\pm p_2}{\sqrt2},\qquad [Q_\alpha,P_\beta]=\ii\hbar\,\delta_{\alpha\beta},
	\label{eq:normal}
\end{equation}
decouples the pair into two independent parametric oscillators,
\begin{equation}
	H(t)=H_+(t)+H_-(t),\qquad
	H_\pm(t)=\frac12\left[P_\pm^2+\Omega_\pm^2(t)\,Q_\pm^2\right],
	\quad \Omega_\pm^2(t)=\omega^2(t)\pm\lambda.
	\label{eq:Hnormal}
\end{equation}
We use \(k\in\{1,2\}\) and \(k\in\{+,-\}\) interchangeably, with
\(\Omega_1\equiv\Omega_+\) and \(\Omega_2\equiv\Omega_-\).  The normal-mode axes
do not rotate in time because the common time dependence multiplies the identity
in the physical-mode potential matrix; the constant coupling contributes the only
off-diagonal term, which the fixed rotation~\eqref{eq:normal} removes once and for
all.  This structural symmetry is precisely what permits two independent
scalar Ermakov equations.  Unequal frequency profiles would in general make the
normal-mode eigenvectors time dependent and reintroduce a residual coupling
between the transformed modes.

\subsection{Ermakov invariants}
\label{sec:ermakov}

For each normal mode let \(\rho_k(t)>0\), \(k\in\{+,-\}\), solve the
Ermakov--Pinney equation
\begin{equation}
	\ddot\rho_k+\Omega_k^2(t)\,\rho_k=\rho_k^{-3}.
	\label{eq:Ermakov}
\end{equation}
Then the Hermitian operator
\begin{equation}
	I_k(t)=\frac12\left[\left(\rho_kP_k-\dot\rho_kQ_k\right)^2
	+\frac{Q_k^2}{\rho_k^2}\right]
	\label{eq:invariant}
\end{equation}
is a Lewis--Riesenfeld invariant of \(H_k\), i.e.\
\(\partial_t I_k+(\ii\hbar)^{-1}[I_k,H_k]=0\).  The cubic term on the right-hand
side of Eq.~\eqref{eq:Ermakov} acts as a centrifugal barrier that keeps
\(\rho_k(t)>0\) for all \(t\), so the invariant is globally defined; the same
solution can be reconstructed from two independent solutions of the associated
linear oscillator through the Pinney formula~\cite{Pinney1950}.  Direct
integration of Eq.~\eqref{eq:Ermakov} is numerically stable for the
positive-definite protocols considered below.

The exact commutator is
\begin{align}
	[I_k,H_k]=\ii\hbar\bigg[&-\rho_k\dot\rho_k\,P_k^2
	+\rho_k\dot\rho_k\,\Omega_k^2\,Q_k^2 \nonumber\\
	&+\frac12\left(\dot\rho_k^2+\rho_k^{-2}-\rho_k^2\Omega_k^2\right)
	(Q_kP_k+P_kQ_k)\bigg].
	\label{eq:commutator}
\end{align}
It vanishes exactly when the invariant is instantaneously aligned with the
Hamiltonian, for instance at \(\rho_k=\Omega_k^{-1/2}\), \(\dot\rho_k=0\).  Away
from that alignment the quadratic off-diagonal terms encode a squeezing of the
invariant frame relative to the instantaneous energy frame; the invariant
condition nonetheless holds exactly because \(\partial_t I_k\) cancels the
commutator.  As emphasized before, \([I_k,H_k]\neq0\) is a signature
of this frame mismatch rather than a dynamical cause in itself.

\subsection{Invariant creation and annihilation operators}
\label{sec:ladder}

Introduce the time-dependent ladder operators
\begin{equation}
	b_k(t)=\frac{1}{\sqrt{2\hbar}}
	\left[\frac{Q_k}{\rho_k}+\ii(\rho_kP_k-\dot\rho_kQ_k)\right],
	\qquad [b_k,b_l^\dagger]=\delta_{kl},
	\label{eq:b}
\end{equation}
in terms of which the invariant is a number operator,
\begin{equation}
	I_k=\hbar\left(b_k^\dagger b_k+\tfrac12\right).
	\label{eq:Inumber}
\end{equation}
Its eigenstates are generated algebraically,
\(|n_k,t\rangle=(b_k^\dagger)^{n}|0_k,t\rangle/\sqrt{n!}\) with
\(b_k|0_k,t\rangle=0\), and their coordinate representation is the chirped
Hermite--Gauss family
\begin{equation}
	\phi_{n,k}(Q,t)=\frac{H_n\!\big(Q/\sqrt{\hbar}\,\rho_k\big)}
	{\sqrt{2^n n!}\,(\pi\hbar\rho_k^2)^{1/4}}
	\exp\!\left[-\frac{Q^2}{2\hbar\rho_k^2}
	+\frac{\ii\dot\rho_k}{2\hbar\rho_k}Q^2\right].
	\label{eq:phi_n}
\end{equation}
The invariant eigenvectors do not solve the Schr\"odinger equation on their own;
the Lewis--Riesenfeld phase supplies the missing time dependence.  The exact
two-mode solution is
\begin{align}
	|\Psi(t)\rangle&=\sum_{n_+,n_-=0}^{\infty}c_{n_+n_-}\,
	e^{\ii\alpha_{n_+n_-}(t)}\,
	|n_+,t\rangle\otimes|n_-,t\rangle,\label{eq:general_solution}\\
	\alpha_{n_+n_-}(t)&=-\sum_{k=\pm}\left(n_k+\tfrac12\right)
	\int_0^t\frac{ds}{\rho_k^2(s)},
	\label{eq:LRphase}
\end{align}
with constants \(c_{n_+n_-}\) fixed by the initial state; a compact algebraic
derivation of Eq.~\eqref{eq:LRphase} is given in
Appendix~\ref{app:phase}.  Equations
\eqref{eq:general_solution}--\eqref{eq:LRphase} solve the full
Schr\"odinger problem, not only its Gaussian sector.  The remainder of the paper
takes the invariant vacuum \(n_+=n_-=0\), for which the state stays Gaussian and
the entanglement dynamics is encoded entirely in second moments.

\subsection{Two inequivalent initial preparations}
\label{sec:initial}

The Ermakov initial data \((\rho_k(0),\dot\rho_k(0))\) specify a physical quantum
state, not an arbitrary numerical convention, and two natural choices must be
kept apart:
\begin{align}
	\text{physical-product input:}\quad
	&\rho_+(0)=\rho_-(0)=\frac{\xi}{\sqrt{\omega_0}},\qquad
	\rho_k(0)\dot\rho_k(0)=\chi,
	\label{eq:productIC}\\
	\text{coupled-ground input:}\quad
	&\rho_k(0)=\frac{\xi}{\sqrt{\Omega_k(0)}},\qquad
	\dot\rho_k(0)=0.
	\label{eq:groundIC}
\end{align}
For the physical-product input~\eqref{eq:productIC} the two normal-mode
covariance blocks coincide at \(t=0\), so the fixed rotation~\eqref{eq:normal}
maps them to a product of two identical physical-mode Gaussians: the state is
genuinely separable.  Here \(\xi>0\) is a common initial width (a local
single-mode squeezing shared by both physical oscillators, \(r_0=\ln\xi\)) and
\(\chi=\rho_k(0)\dot\rho_k(0)\) is a common dimensionless quadratic-phase chirp;
at \(\xi=1,\chi=0\) the state is the product of the uncoupled \(\omega_0\) vacua.
The coupled-ground input~\eqref{eq:groundIC} with \(\xi=1\) is instead the ground
state of the coupled Hamiltonian and, because \(\Omega_+(0)\neq\Omega_-(0)\) for
\(\lambda\neq0\), is already entangled in the \((x_1,x_2)\) partition.  We
 use the physical-product input for every claim of entanglement
generation and retain the coupled-ground state only as an adiabatic benchmark.

\section{Gaussian covariance matrix and entanglement}
\label{sec:gaussian}

\subsection{Covariance matrix}
\label{sec:cm}

Quadratic Hamiltonians preserve Gaussianity, so first and second moments give a
complete description of the states considered here~\cite{Weedbrook2012}.  With
zero first moments, the state is fixed by the covariance matrix
\(\CM_{ij}=\tfrac12\langle\{\Delta R_i,\Delta R_j\}\rangle\).  For the invariant
vacuum, Gaussian integration of Eq.~\eqref{eq:phi_n} with \(n=0\)
(Appendix~\ref{app:cov}) gives the normal-mode block in the ordering
\((Q_k,P_k)\),
\begin{equation}
	\CM_k(t)=\frac{\hbar}{2}
	\begin{pmatrix}
		\rho_k^2&\rho_k\dot\rho_k\\[2pt]
		\rho_k\dot\rho_k&\dot\rho_k^2+\rho_k^{-2}
	\end{pmatrix},
	\qquad \det\CM_k=\left(\frac{\hbar}{2}\right)^2,
	\label{eq:modeCM}
\end{equation}
so each mode, and hence the global state, is pure.  Transforming
\(\CM^{\rm nm}=\CM_+\oplus\CM_-\) back to the physical ordering
\(R=(x_1,p_1,x_2,p_2)^{\trans}\) with the symplectic map induced
by~\eqref{eq:normal} yields the standard two-mode block form
\begin{equation}
	\CM=\begin{pmatrix}A&C\\C^{\trans}&B\end{pmatrix},\qquad
	A=B=\frac{\CM_++\CM_-}{2},\quad
	C=\frac{\CM_+-\CM_-}{2}.
	\label{eq:physicalCM}
\end{equation}
The equality \(A=B\) reflects the exchange symmetry \(\omega_1=\omega_2\).  A
nonzero \(C\) signals correlations but is not by itself an entanglement test;
for the present globally pure state, however, \(\CM_+\neq\CM_-\) already makes
the single-mode reduced state mixed, which for a pure bipartite state is
equivalent to entanglement.

\subsection{Closed logarithmic negativity}
\label{sec:negativity}

Partial transposition of mode~2 is the phase-space reflection
\(p_2\mapsto-p_2\)~\cite{Simon2000}, under which the smallest symplectic
eigenvalue of the transposed covariance matrix is
\begin{equation}
	\widetilde\nu_-^2=\frac{\widetilde\Delta-
		\sqrt{\widetilde\Delta^2-4\det\CM}}{2},\qquad
	\widetilde\Delta=\det A+\det B-2\det C.
	\label{eq:nupt}
\end{equation}
The logarithmic negativity  is~\cite{Plenio2005,VidalWerner2002}, which was shown to be an entanglement monotone in~\cite{Plenio2005}
\begin{equation}
	\EN=\max\left\{0,\,-\ln\frac{2\widetilde\nu_-}{\hbar}\right\}.
	\label{eq:ENdef}
\end{equation}
For the diagonal blocks~\eqref{eq:modeCM} all dependence collapses onto a single
dimensionless invariant,
\begin{equation}
	X(t)=\left(\rho_+\dot\rho_--\rho_-\dot\rho_+\right)^2
	+\left(\frac{\rho_+}{\rho_-}\right)^2
	+\left(\frac{\rho_-}{\rho_+}\right)^2\ \geq\ 2,
	\label{eq:X}
\end{equation}
and a short computation (Appendix~\ref{app:negativity}) gives
\begin{equation}
	\left(\frac{2\widetilde\nu_-}{\hbar}\right)^2
	=\frac{X-\sqrt{X^2-4}}{2},\qquad
	\ \EN(t)=\frac12\operatorname{arcosh}\frac{X(t)}{2}\ .
	\label{eq:closedEN}
\end{equation}
The bound \(X\geq2\) is saturated, and hence \(\EN=0\), precisely when
\(\rho_+=\rho_-\) and \(\dot\rho_+=\dot\rho_-\) simultaneously.  The amplitude
ratios in Eq.~\eqref{eq:X} measure a differential width and the Wronskian-like
term \(W\equiv\rho_+\dot\rho_--\rho_-\dot\rho_+\) a differential phase-space
shear; entanglement is therefore produced by the relative Gaussian
deformation of the two normal modes, not by the amount of squeezing in either
one alone.\footnote{Near threshold \(X-2\simeq W^2+(\rho_+/\rho_--\rho_-/\rho_+)^2\);
	since \(W\) and the amplitude mismatch both vanish as \(\lambda\to0\), one has
	\(\EN\simeq\tfrac12\sqrt{X-2}\), linear in the coupling.}

\subsection{Logarithmic negativity versus the Duan--Simon witness}
\label{sec:duan}

The Duan criterion uses the EPR-like quadratures~\cite{Duan2000}
\begin{equation}
	u=|a|\,\bar x_1+a^{-1}\bar x_2,\qquad
	v=|a|\,\bar p_1-a^{-1}\bar p_2,
	\label{eq:EPR}
\end{equation}
for which every separable state obeys
\begin{equation}
	\mathcal D(a)\equiv\operatorname{Var}u+\operatorname{Var}v
	\ \geq\ \hbar\,(a^2+a^{-2}).
	\label{eq:Duan}
\end{equation}
The bars are essential: the physical covariance matrix~\eqref{eq:physicalCM} is
generally not in standard form, because the off-diagonal block
\(C=\tfrac12(\CM_+-\CM_-)\) inherits the \(\rho_k\dot\rho_k\) entries.  A fair
application therefore requires first bringing \(\CM\) to standard form by local
symplectic (single-mode) operations, which leave the entanglement unchanged, and
then optimizing over \(a\).  For the globally pure, exchange-symmetric Gaussian
family studied here this local standardization maps the state to a two-mode
squeezed vacuum, and the optimized Duan ratio becomes
\begin{equation}
	\ \frac{\mathcal D_{\rm opt}}{\mathcal D_{\rm sep}}
		=\frac{2\widetilde\nu_-}{\hbar}=e^{-\EN}\ ,
	\label{eq:DuanRelation}
\end{equation}
so that \(\mathcal D_{\rm opt}<\mathcal D_{\rm sep}\) if and only if \(\EN>0\):
the Duan witness and the positive-partial-transpose criterion detect
entanglement at the same threshold and order this restricted family
identically (Appendix~\ref{app:duan}).  A convenient bounded score is
\begin{equation}
	W_D=\max\left\{0,\,1-\mathcal D_{\rm opt}/\mathcal D_{\rm sep}\right\}
	=1-e^{-\EN}.
	\label{eq:DuanScore}
\end{equation}
The equivalence is special to this pure symmetric family: outside it a raw or
even optimized Duan variance is not a one-to-one function of \(\EN\), so \(W_D\)
is best read as an optimized inseparability witness rather than as an
independent entanglement measure.  We therefore adopt \(\EN\) as the primary
measure and use the Duan score only as the experimentally economical witness,
which requires two optimized quadrature variances rather than the full
covariance matrix.

\section{Sinusoidal modulation}
\label{sec:numerics}

As an experimentally relevant illustration we modulate the common
stiffness rather than the frequency itself,
\begin{equation}
	\omega^2(t)=\omega_0^2\bigl[1+\epsilon\sin(\nu t)\bigr],\qquad
	\Omega_\pm^2(t)=\omega_0^2\bigl[1+\epsilon\sin(\nu t)\bigr]\pm\lambda,
	\label{eq:drive}
\end{equation}
with \(0<\epsilon<1-|\lambda|/\omega_0^2\) to keep the potential positive
definite.  Modulating \(\omega^2\) linearly in \(\sin\nu t\) yields the standard
Mathieu-type parametric drive without the spurious second harmonic that would
arise from squaring \(1+\epsilon\sin\nu t\).  Unless otherwise stated we use the
dimensionless parameters
\begin{equation}
	\omega_0=1,\qquad \epsilon=0.25,\qquad \lambda=0.12,\qquad \xi=1,\quad\chi=0,
	\label{eq:parameters}
\end{equation}
and the separable physical-product preparation~\eqref{eq:productIC}.  The two
Ermakov equations and Lewis--Riesenfeld phase integrals are propagated with an
eighth-order explicit Runge--Kutta scheme (relative tolerance \(2\times10^{-10}\),
absolute tolerance \(2\times10^{-12}\)); throughout the evolution the closed
expression~\eqref{eq:closedEN} agrees with the smallest symplectic eigenvalue of
the full \(4\times4\) covariance matrix to better than \(3\times10^{-13}\).

\begin{figure}[t]
	\centering
	\includegraphics[width=\textwidth]{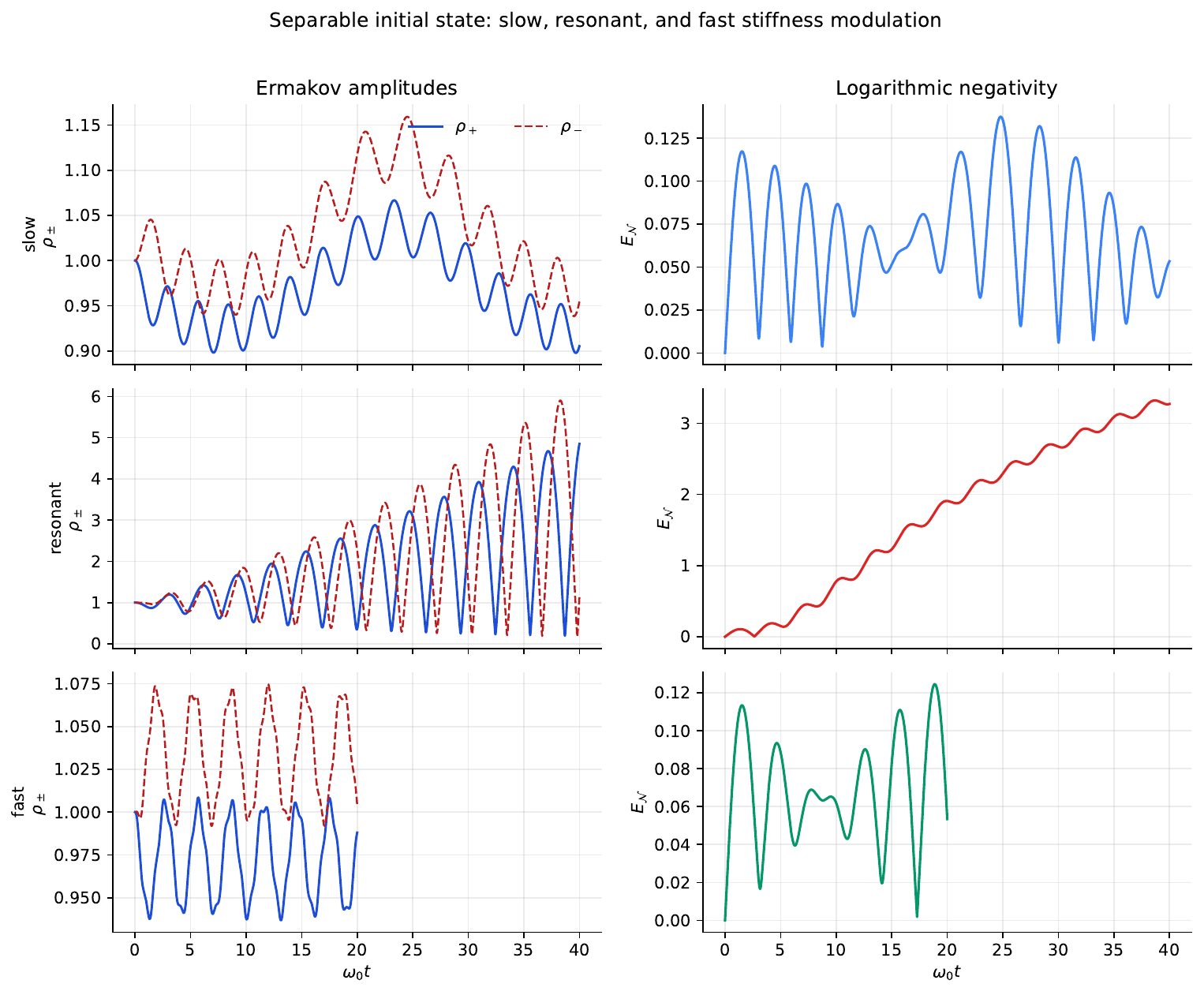}
	\caption{Ermakov amplitudes \(\rho_\pm(t)\) (left) and logarithmic negativity
		\(\EN(t)\) (right) for the separable physical-product input under slow,
		near-resonant, and fast sinusoidal stiffness modulation.  Parameters are those
		of Eq.~\eqref{eq:parameters}; the fast-drive window is shortened for
		visibility.}
	\label{fig:regimes}
\end{figure}

Figure~\ref{fig:regimes} contrasts three driving regimes: slow
(\(\nu=0.2\,\omega_0\)), near-resonant (\(\nu=2\,\omega_0\)), and fast
(\(\nu=8\,\omega_0\)).  The maximum logarithmic negativities over the displayed
windows are \(0.138\), \(3.321\), and \(0.125\), respectively.  Slow and fast
driving produce bounded beating in which \(\rho_\pm\) oscillate near unity and
\(\EN\) stays small.  Near twice the mean normal-mode frequency, parametric
amplification drives the two Ermakov trajectories progressively apart and
\(\EN\) grows through successive steps.  Because \(\Omega_+\neq\Omega_-\), the
two modes have slightly displaced resonance bands and respond to the drive at
slightly different rates; this differential response is exactly what the
invariant \(X(t)\) of Eq.~\eqref{eq:X} quantifies.

\begin{figure}[t]
	\centering
	\includegraphics[width=.82\textwidth]{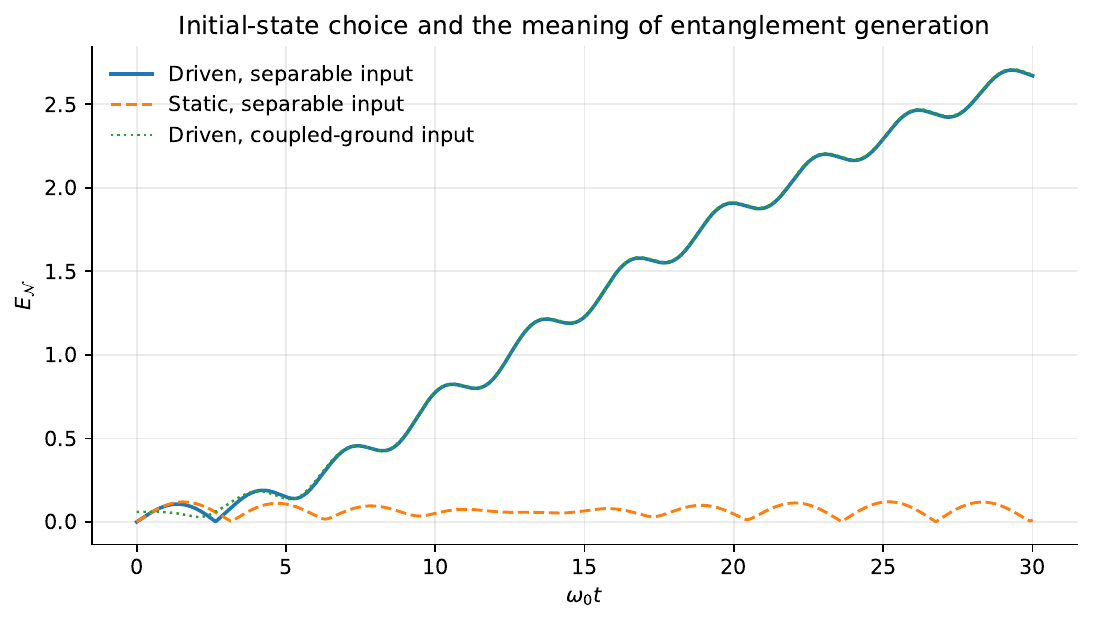}
	\caption{Driven and undriven dynamics for the separable physical-product input,
		together with the driven coupled-ground benchmark.  The benchmark is entangled
		already at \(t=0\), whereas the product input starts at \(\EN=0\).}
	\label{fig:baseline}
\end{figure}

Modulation is not the only source of entanglement: the constant coupling alone
also evolves a physical-product vacuum into an entangled state.  To isolate the
role of the drive, Figure~\ref{fig:baseline} adds the undriven (\(\epsilon=0\))
baseline, which stays below \(\EN\simeq0.12\) over the plotted window, whereas
the resonant drive produces sustained growth.  The driven coupled-ground
benchmark starts at \(\EN(0)=0.0603\), confirming that it is already
entangled and cannot serve as a separable initial condition.

\begin{figure}[t]
	\centering
	\includegraphics[width=.96\textwidth]{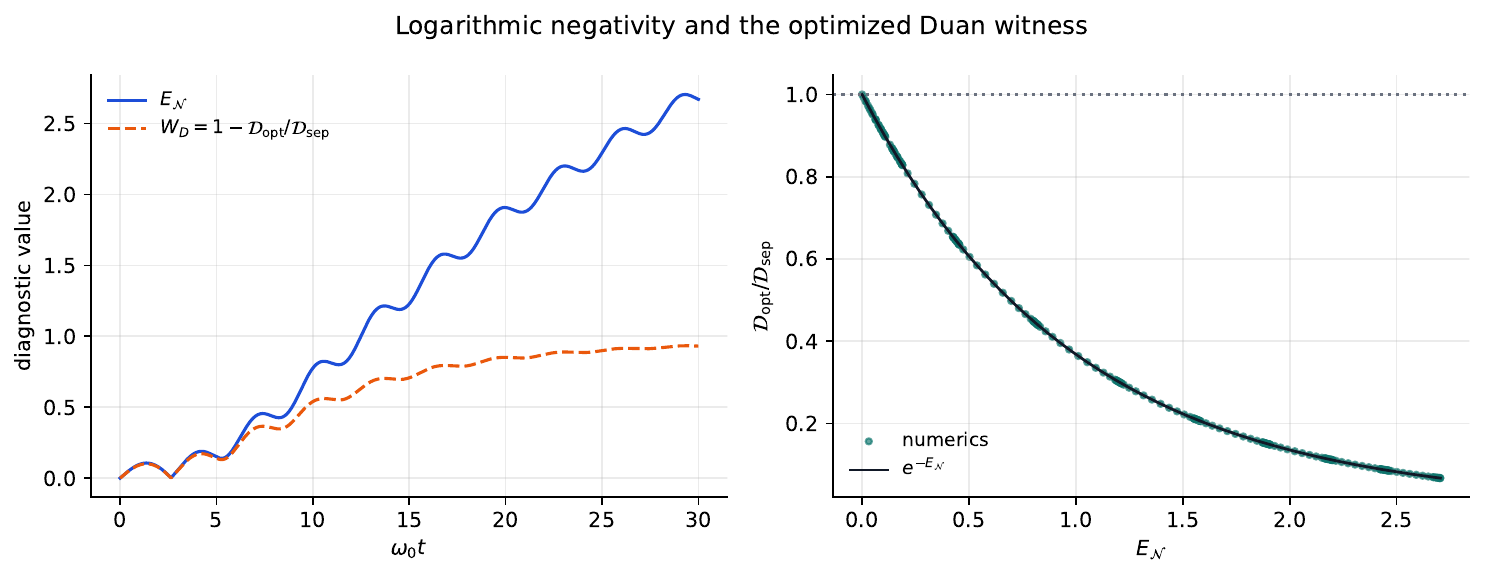}
	\caption{Left: logarithmic negativity \(\EN(t)\) and the bounded optimized Duan
		witness score \(W_D=1-\mathcal D_{\rm opt}/\mathcal D_{\rm sep}\).  Right:
		direct numerical confirmation that
		\(\mathcal D_{\rm opt}/\mathcal D_{\rm sep}=e^{-\EN}\) for the present state
		family.}
	\label{fig:duan}
\end{figure}

Figure~\ref{fig:duan} verifies Eq.~\eqref{eq:DuanRelation} numerically: the
optimized Duan witness and the logarithmic negativity cross the separability
threshold together, while their values are related by the nonlinear map
\(\mathcal D_{\rm opt}/\mathcal D_{\rm sep}=e^{-\EN}\).  Experimental use of the
witness still requires either prior knowledge of the appropriate local
quadratures or an explicit optimization; two unoptimized laboratory variances
need not saturate Eq.~\eqref{eq:DuanRelation}.

\section{Control through the initial Gaussian state}
\label{sec:control}

For a fixed frequency protocol the entire evolution is determined by the initial
Ermakov data, so the target-time entanglement is a deterministic function of the
two preparation parameters of Eq.~\eqref{eq:productIC},
\begin{equation}
	J(\xi,\chi;T)=\EN\bigl(T\mid\xi,\chi\bigr).
	\label{eq:objective}
\end{equation}
Both controls act before the evolution and preserve separability of the
input: \(\xi\) is a common width squeezing shared by the two physical
oscillators (\(r_0=\ln\xi\)) and \(\chi=\rho_k(0)\dot\rho_k(0)\) a common
quadratic phase, so the two normal-mode blocks remain identical at \(t=0\) for
every \((\xi,\chi)\).  This is what lets us attribute the entire final
entanglement to controlled generation rather than to initial correlations.

\begin{figure}[t]
	\centering
	\includegraphics[width=\textwidth]{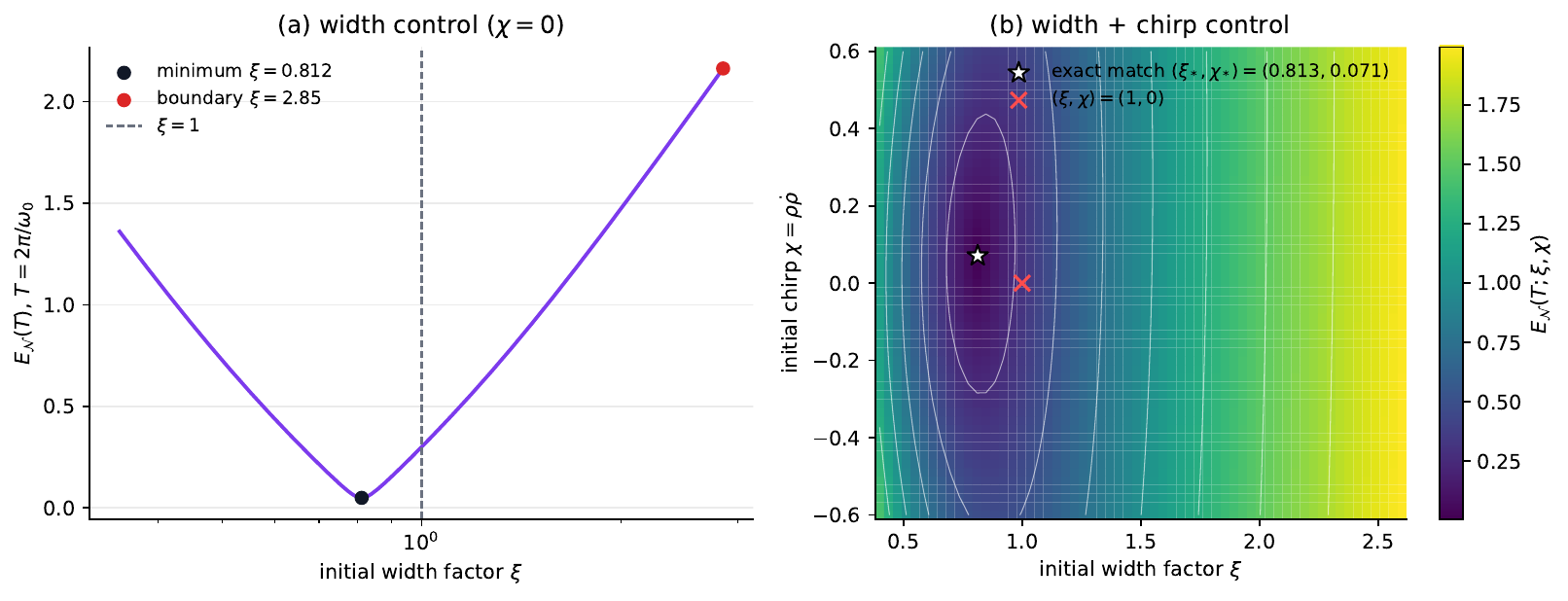}
	\caption{Target logarithmic negativity at \(T=2\pi/\omega_0\).  (a)~Width control
		along \(\chi=0\): an interior minimum near \(\xi=0.812\) and monotone growth to
		the boundaries.  (b)~Width\,\(+\,\)chirp control: the two-parameter landscape
		\(J(\xi,\chi;T)\).  The minimizing valley is tilted, so the optimum is displaced
		from the unsqueezed input \((\xi,\chi)=(1,0)\) (cross).  The star marks the
		numerically solved matching point
		\((\xi_*,\chi_*)=(0.81279548,0.07096760)\), where both conditions in
		Eq.~\eqref{eq:matching} vanish and \(\EN(T)=0\) to numerical precision.}
	\label{fig:control}
\end{figure}

Figure~\ref{fig:control}(a) shows the one-dimensional slice \(\chi=0\) at
\(T=4\pi/\nu=2\pi/\omega_0\) (two drive periods).  On the accessible interval
\(0.35\leq\xi\leq2.85\) the target entanglement has a shallow interior minimum
\(\EN=0.0496\) at \(\xi=0.8118\), rises to \(\EN=0.3004\) at the unsqueezed point
\(\xi=1\), and reaches its largest values at the boundaries,
\(\EN=1.3604\) at \(\xi=0.35\) and \(\EN=2.1630\) at \(\xi=2.85\).  The landscape
is manifestly not a function of \(|\ln\xi|\): squeezing the input by the same
factor in width does not produce the same entanglement, and there is no universal
interior optimum independent of \(T,\nu,\epsilon,\lambda\).  Within experimentally
imposed squeezing bounds the optimal strategy is therefore to squeeze the input
as strongly as the platform allows, away from the interior minimum.

Adding the chirp \(\chi\) opens a genuinely two-dimensional control plane,
Figure~\ref{fig:control}(b).  The entanglement-minimizing valley is tilted in
\((\xi,\chi)\), so the true minimum is displaced from the unsqueezed point
\((1,0)\).  We determine the zero directly, rather than inferring it from the
plotting grid, by solving the two target-time equations
\begin{equation}
 F_1(\xi,\chi)=\rho_+(T)-\rho_-(T)=0,\qquad
 F_2(\xi,\chi)=\dot\rho_+(T)-\dot\rho_-(T)=0.
 \label{eq:matching}
\end{equation}
For the parameters of Fig.~\ref{fig:control}, a nonlinear root solve gives
\((\xi_*,\chi_*)=(0.81279548,0.07096760)\), with
\(\|\bm F(\xi_*,\chi_*)\|_\infty<4\times10^{-14}\).  At this point
\(\rho_+(T)=\rho_-(T)=1.22836055\) and
\(\dot\rho_+(T)=\dot\rho_-(T)=0.10439033\), hence \(X(T)-2\) vanishes to
double precision and both the closed formula and the general covariance-matrix
calculation give \(\EN(T)<2\times10^{-15}\).  Thus the two controls null the
target-time entanglement, whereas width control alone cannot fall below
\(0.0496\).  The chirp supplies precisely the second degree of freedom needed
to enforce Eq.~\eqref{eq:matching}.

A practical open-loop protocol follows directly:
\begin{enumerate}[leftmargin=2em]
	\item calibrate \(\omega^2(t)\), \(\lambda\), and the target time \(T\);
	\item determine the experimentally allowed rectangle in \((\xi,\chi)\);
	\item integrate Eq.~\eqref{eq:Ermakov} and optimize Eq.~\eqref{eq:objective}
	over that rectangle;
	\item prepare identical local Gaussian states of the two physical oscillators;
	\item apply the calibrated modulation and verify the target with optimized EPR
	quadratures or full covariance tomography.
\end{enumerate}
All control effort is invested in the initial-state preparation; no
time-dependent feedback is required during the evolution.  The present unitary
calculation does not include thermal loss, dephasing, or control-amplitude
noise, so robustness figures should not be quoted without specifying a master
equation and a noise spectrum; a first estimate of the coherence budget for a
representative platform is given in Sec.~\ref{sec:discussion}, and a full
open-system treatment is left to future work.

\section{\texorpdfstring{$SU(1,1)$}{SU(1,1)} interpretation}
\label{sec:su11}

The dynamics has a compact geometric reading.  Fix a reference frequency
\(\Omega_r>0\) with ladder operator \(a_k\), and define the \(SU(1,1)\) generators
\begin{equation}
	K_0^{(k)}=\tfrac12\left(a_k^\dagger a_k+\tfrac12\right),\qquad
	K_+^{(k)}=\tfrac12 a_k^{\dagger2},\qquad
	K_-^{(k)}=\tfrac12 a_k^2 .
	\label{eq:generators}
\end{equation}
In this frame the normal-mode Hamiltonian is
\begin{equation}
	H_k(t)=\hbar\,\frac{\Omega_r^2+\Omega_k^2(t)}{\Omega_r}\,K_0^{(k)}
	+\hbar\,\frac{\Omega_k^2(t)-\Omega_r^2}{2\Omega_r}
	\left(K_+^{(k)}+K_-^{(k)}\right),
	\label{eq:suH}
\end{equation}
so the \(K_\pm\) terms are algebraic squeezing generators that switch on whenever
\(\Omega_k(t)\neq\Omega_r\); the propagator is an \(SU(1,1)\) element, a squeeze
followed by a rotation.  For the pure covariance~\eqref{eq:modeCM} the squeeze
magnitude and principal-axis angle relative to \(\Omega_r\) are
\begin{align}
	r_k&=\tfrac12\operatorname{arcosh}\!\left[\tfrac12\!\left(
	\Omega_r\rho_k^2+\frac{\dot\rho_k^2+\rho_k^{-2}}{\Omega_r}\right)\right],
	\label{eq:r}\\[2pt]
	2\theta_k&=\atantwo\!\left(2\rho_k\dot\rho_k,\;
	\Omega_r\rho_k^2-\frac{\dot\rho_k^2+\rho_k^{-2}}{\Omega_r}\right).
	\label{eq:theta}
\end{align}
Writing \(M_k=2\CM_k/\hbar\) in the dimensionless reference frame (unit
determinant), the invariant of Sec.~\ref{sec:negativity} is a single
\(SU(1,1)\) trace,
\begin{align}
	X&=\tr\!\left(M_+^{-1}M_-\right)\nonumber\\
	&=2\Bigl[\cosh(2r_+)\cosh(2r_-)
	-\cos\!\bigl(2\theta_+-2\theta_-\bigr)\sinh(2r_+)\sinh(2r_-)\Bigr].
	\label{eq:Xsu}
\end{align}
Equation~\eqref{eq:Xsu} makes the geometric content explicit: with
\(d_{\rm hyp}=\operatorname{arcosh}(X/2)\) the hyperbolic separation of the two
mode points on the \(SU(1,1)/U(1)\) disk,
\begin{equation}
	\ \EN=\tfrac12\,d_{\rm hyp}\ .
	\label{eq:hyperbolic}
\end{equation}

\begin{figure}[t]
	\centering
	\includegraphics[width=\textwidth]{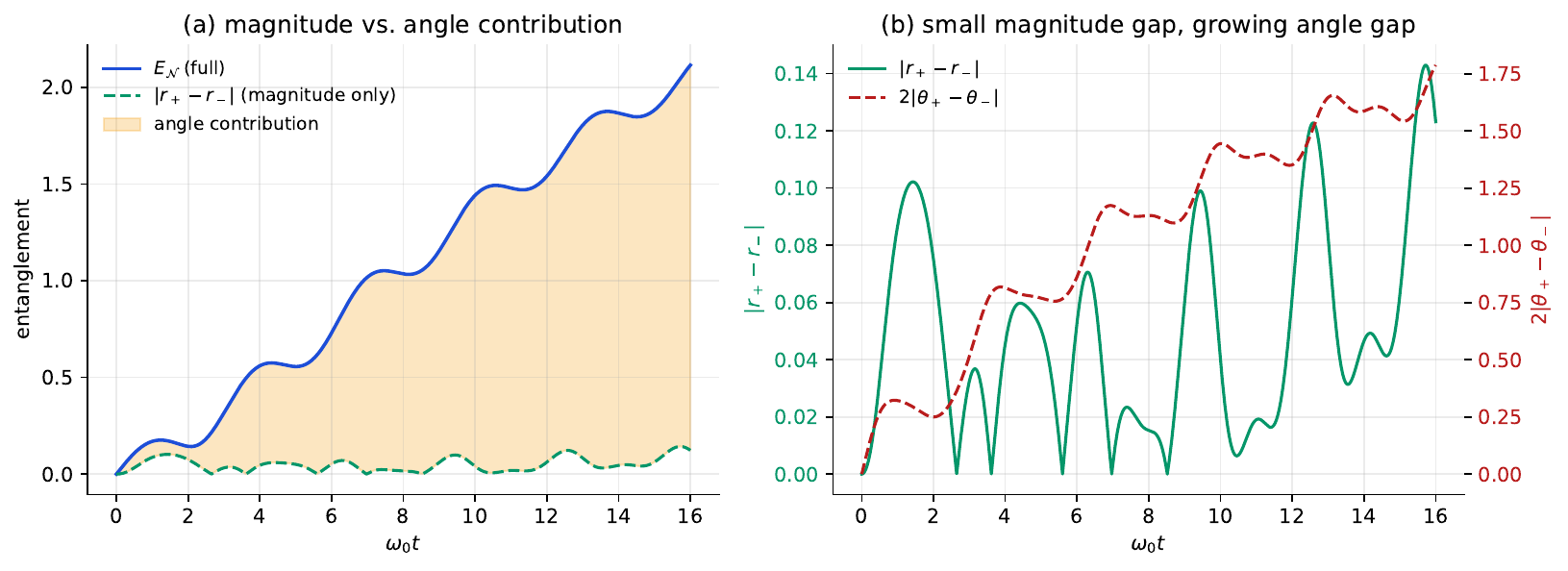}
	\caption{Angle-driven entanglement.  (a)~The full logarithmic negativity
		\(\EN(t)\) (solid) versus the aligned-axes, magnitude-only prediction
		\(|r_+-r_-|\) (dashed); the shaded gap is the relative-angle contribution of
		Eq.~\eqref{eq:Xsu}.  (b)~The magnitude gap \(|r_+-r_-|\) stays small while the
		relative angle \(2|\theta_+-\theta_-|\) grows, so the entanglement is carried by
		the angle.  Separable input with \(\xi=1.4\), parameters~\eqref{eq:parameters}.}
	\label{fig:angle}
\end{figure}

Crucially, Eq.~\eqref{eq:Xsu} corrects the oversimplified statement that
entanglement is controlled by the magnitude difference \(|r_+-r_-|\) alone.  Only
when the squeezing axes are aligned (\(\theta_+=\theta_-\)) does
Eq.~\eqref{eq:Xsu} reduce to \(X=2\cosh\!\bigl(2(r_+-r_-)\bigr)\) and hence
\(\EN=|r_+-r_-|\); in general the relative angle contributes on an equal
footing, and can generate strong entanglement even when the two magnitudes
coincide.  Figure~\ref{fig:angle} demonstrates this in the driven dynamics: the
full \(\EN(t)\) greatly exceeds the aligned-axes prediction \(|r_+-r_-|\), the gap
(shaded) being the angle contribution.  At \(\omega_0 t\approx8.6\), for example,
the two modes are squeezed by nearly equal amounts
(\(r_+\approx r_-\approx0.81\), \(|r_+-r_-|\approx0.005\)) yet
\(\EN\approx1.06\), driven almost entirely by the relative angle
\(2|\theta_+-\theta_-|\approx1.1\).  A purely synthetic check confirms the
mechanism: two modes with equal magnitude \(r=0.6\) and orthogonal axes
(\(\theta_+-\theta_-=\pi/2\)) give \(\EN=1.20\), while aligned axes give
\(\EN=0\).

\begin{figure}[t]
	\centering
	\includegraphics[width=.57\textwidth]{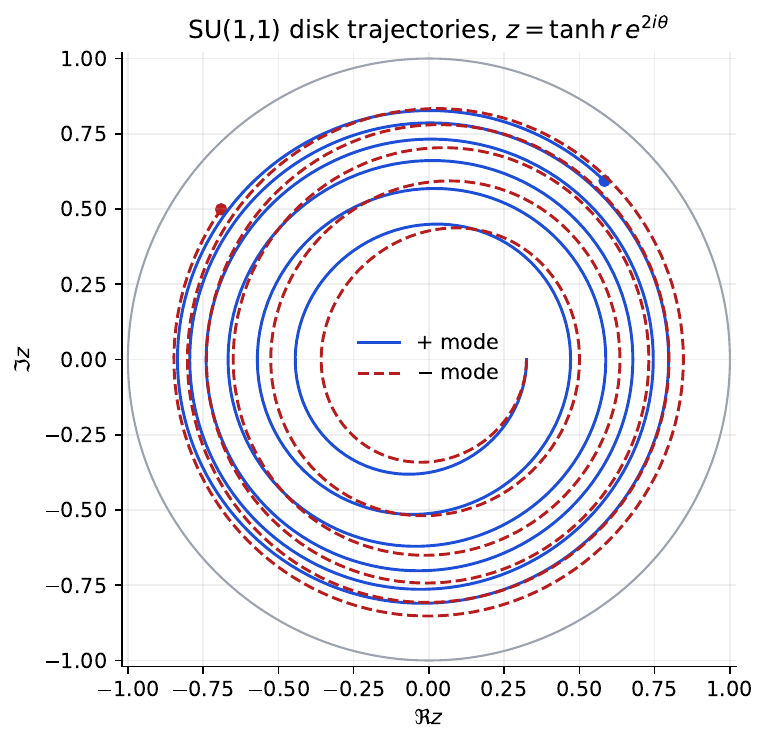}
	\caption{Normal-mode trajectories on the \(SU(1,1)\) disk,
		\(z=\tanh r\,e^{2\ii\theta}\), for a separable input with \(\xi=1.4\).  It is
		their hyperbolic separation, not the Euclidean separation in the drawing, that
		fixes the logarithmic negativity through Eq.~\eqref{eq:hyperbolic}.}
	\label{fig:disk}
\end{figure}

Geometrically, the initial width and chirp place both modes at a common starting
point on the disk, while the split frequency profiles \(\Omega_\pm(t)\) send them
along two trajectories whose hyperbolic separation at the target time is the
control objective (Fig.~\ref{fig:disk}).  Maximizing entanglement at \(T\) means
choosing the initial point so that the two trajectories end as far apart as
possible in the hyperbolic metric.  This viewpoint suggests further control
strategies beyond the present initial-state protocol, including shaping
\(\Omega(t)\) to steer the trajectories along geodesics of the \(SU(1,1)/U(1)\)
manifold and multi-tone driving to engineer composite trajectories.

\section{Discussion, limitations, and scalability}
\label{sec:discussion}

\emph{Scope of the exact result.\\}
The closed-form treatment rests on four assumptions: equal masses, identical
physical frequency profiles, constant bilinear coupling, and an initial pure
Gaussian state that is a product in the normal-mode basis.  The first three keep
the normal-mode transformation fixed in time.  A time-dependent coupling
\(\lambda\to\lambda(t)\) can be accommodated without changing the eigenvectors as
long as the potential retains the exchange symmetry, since only the eigenvalues
\(\Omega_\pm^2(t)=\omega^2(t)\pm\lambda(t)\) then move.  Unequal local profiles,
by contrast, make the normal-mode eigenvectors time dependent and reintroduce a
residual coupling between the transformed modes, so two independent scalar
Ermakov equations no longer close the problem; that regime requires a genuine
time-dependent Bogoliubov rotation and is beyond the present scope.

\begin{figure}[t]
	\centering
	\includegraphics[width=.72\textwidth]{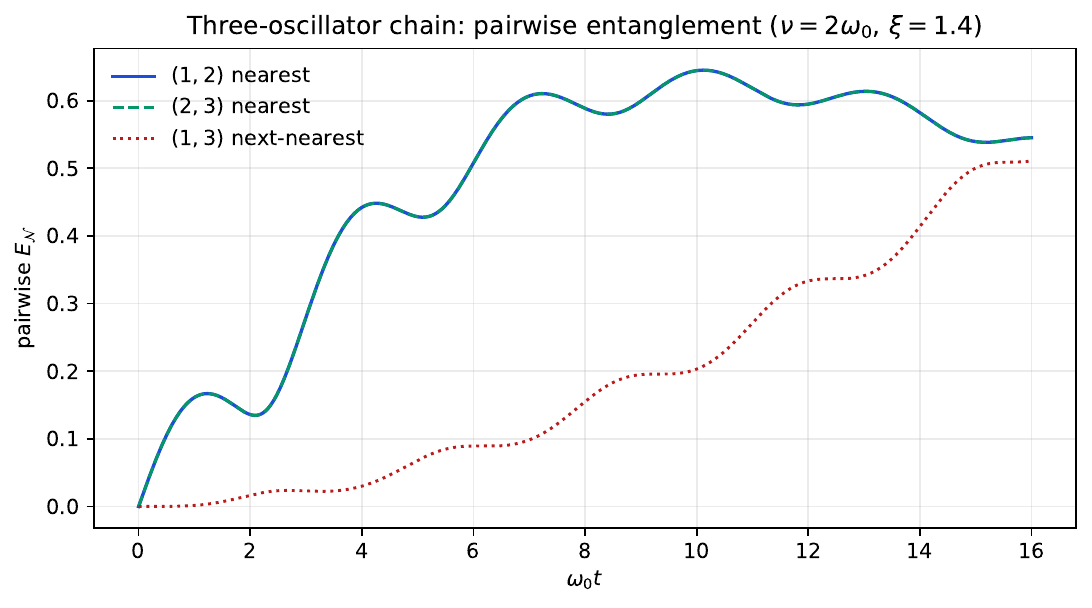}
	\caption{Scalability example: pairwise logarithmic negativities in a
		three-oscillator chain with nearest-neighbour coupling
		(\(V=\omega^2(t)\mathbb{I}+K\)), common resonant modulation
		(\(\nu=2\omega_0\)), and a separable physical-product input (\(\xi=1.4\)).  The
		two nearest-neighbour pairs coincide by symmetry; the next-nearest pair is
		generated more weakly.}
	\label{fig:network}
\end{figure}

\emph{Scalability to networks.\\}
For \(N\) oscillators with a potential \(V(t)=\omega^2(t)\,\mathbb{I}+K\) and a
constant coupling matrix \(K\), a single time-independent orthogonal
transformation \(O\) diagonalizes \(K\), giving \(N\) decoupled parametric
oscillators with stiffnesses \(\Omega_a^2(t)=\omega^2(t)+\mu_a\) (\(\mu_a\) the
eigenvalues of \(K\)).  One then integrates \(N\) independent Ermakov equations
--- an \(O(N N_t)\) task after the one-time \(O(N^3)\) diagonalization --- and
builds the physical covariance matrix by the fixed congruence
\(\CM=S(\bigoplus_a\CM_a)S^{\trans}\), with \(S\) induced by \(O\).  Any two-site
Gaussian entanglement diagnostic then follows from a \(4\times4\) reduced block: for two physical
oscillators \((i,j)\) one keeps the corresponding rows and columns of \(\CM\) and
evaluates the general (mixed-state) logarithmic negativity of
Eq.~\eqref{eq:nupt}.  Figure~\ref{fig:network} shows this for a three-oscillator
chain with nearest-neighbour coupling under resonant modulation of a separable
input: all pairwise negativities start at zero, the two nearest-neighbour pairs
(equal by the chain symmetry) grow to \(\EN\approx0.65\), and the next-nearest
pair to \(\EN\approx0.51\).  Because the global three-mode state is pure and
every one-mode-versus-two-mode cut has a nonzero pairwise-entangled reduction,
it cannot factorize across any such cut; common modulation and fixed coupling
therefore generate genuine tripartite Gaussian entanglement.  The dynamical cost is therefore dominated by covariance
manipulation, not by the ODE integration.

\emph{Experimental mapping and coherence budget.\\}
The dimensionless example establishes a mechanism, not a platform-specific
feasibility claim, but it maps naturally onto existing hardware.  Table~\ref{tab:ion}
translates the parameters to a representative trapped-ion realization with a
secular frequency \(\omega_0/2\pi=1\,\text{MHz}\)~\cite{Leibfried2003}: the target
time is \(T=2\pi/\omega_0\approx1\,\mu\text{s}\), the resonant modulation is at
\(2\,\text{MHz}\) with a \(25\%\) stiffness depth, and the normal-mode separation is
\(\Omega_+-\Omega_-=0.1202\,\omega_0\approx2\pi\times120\,\text{kHz}\).  The often
quoted \(60\,\text{kHz}\) scale is instead the approximate displacement of each
normal mode from \(\omega_0\) (about \(+58\) and \(-62\,\text{kHz}\)).  A nine-decibel
reduction of a motional quadrature variance has been demonstrated directly in a
trapped ion~\cite{Lo2015}.  Experiments have also reported motional heating near
two phonons per second and coherence beyond \(8\,\text{ms}\)~\cite{Maslennikov2019};
a recent room-temperature device reaches \(1.1(1)\) quanta/s and a motional
Ramsey coherence time near \(96\,\text{ms}\)~\cite{Menon2026}.  Using the
conservative illustrative values \(\tau=8\,\text{ms}\) and
\(\dot{\bar n}=2\,\text{s}^{-1}\) gives \(T/\tau\approx1.3\times10^{-4}\) and
\(\Delta\bar n\approx2\times10^{-6}\) during the target interval.  These separate
experimental benchmarks support the order-of-magnitude coherence budget; they do
not establish simultaneous implementation of every entry in Table~\ref{tab:ion},
and a platform-specific robustness claim still requires a master equation with
the measured noise spectrum.

\begin{table}[t]
	\centering
	\small
	\begin{tabular}{@{}lll@{}}
		\toprule
		Quantity & Dimensionless & Trapped-ion value \\
		\midrule
		Secular frequency & \(\omega_0\) & \(2\pi\times1\,\text{MHz}\) \\
		Modulation frequency & \(\nu=2\omega_0\) & \(2\pi\times2\,\text{MHz}\) \\
		Target time & \(T=2\pi/\omega_0\) & \(\approx1\,\mu\text{s}\) \\
		Modulation depth & \(\epsilon=0.25\) & \(25\%\) of \(\omega_0^2\) \\
		Mode splitting & \(\lambda=0.12\,\omega_0^2\) & \(\Omega_+-\Omega_-\approx2\pi\times120\,\text{kHz}\) \\
		Width control & \(r_0=\ln\xi\in[-1.05,1.05]\) & up to \(\approx9\,\text{dB}\) squeezing \\
		Coherence budget & \(T/\tau\sim10^{-4}\) & \(\tau=8\,\text{ms}\), \(\Delta\bar n\sim2\times10^{-6}\) \\
		\bottomrule
	\end{tabular}
	\caption{Representative mapping of the dimensionless parameters to a trapped-ion
		platform.  The control window is short compared with motional coherence, so the
		unitary result applies over the target time.}
	\label{tab:ion}
\end{table}

\emph{Inverse engineering.\\}
Finally, the invariant formalism is naturally suited to inverse
engineering~\cite{Tobalina2020}: rather than optimizing only the initial state,
one may prescribe boundary conditions on \(\rho_k(t)\) and reconstruct a
permissible frequency profile \(\omega^2(t)\).  A complete treatment would impose
bandwidth and positivity constraints from the outset, and combining shaped
\(\Omega(t)\) with initial-state control is a promising route to
shortcut-to-adiabaticity generation of a prescribed entanglement at \(T\).

\section{Conclusions}
\label{sec:conclusions}

We have given an exact, closed-form, and fully reproducible description of
Gaussian entanglement in two symmetrically coupled oscillators with a common
time-dependent stiffness.  The Lewis--Riesenfeld invariant solves the
Schr\"odinger equation algebraically, and its Ermakov amplitudes deliver the
physical covariance matrix directly, so the entire bipartite entanglement is
governed by a single relative invariant \(X(t)\) through
\(\EN=\tfrac12\operatorname{arcosh}(X/2)\).  For the pure exchange-symmetric
family the optimized Duan witness detects the same threshold with the
normalization-explicit relation
\(\mathcal D_{\rm opt}/\mathcal D_{\rm sep}=e^{-\EN}\), which we stress is a
witness rather than an independent measure.

Three points give the analysis its physical content.  First, the distinction
between a genuinely separable physical-product input and the already-entangled
ground state of the coupled Hamiltonian is essential to any claim of entanglement
generation, and we use only the former.  Second, the initial width and
chirp are legitimate, experimentally accessible controls: they reshape the input
Gaussian state without inserting correlations, and together they span a
two-parameter landscape that can steer the target-time entanglement across its
full range --- from an exact separable point to the boundary of the accessible
squeezing.  Third, the \(SU(1,1)\) formulation identifies \(2\EN\) with a
hyperbolic distance on the \(SU(1,1)/U(1)\) disk and shows that entanglement is
set by both the relative squeeze magnitude and the relative squeeze angle; in
particular, nearly equal magnitudes can still entangle strongly through the angle
alone.

The framework extends without modification to \(N\) oscillators that share a
common modulation and a fixed coupling topology, where a single diagonalization
plus \(N\) scalar Ermakov equations generate multipartite Gaussian entanglement
from a separable input.  Natural continuations include an open-system robustness
analysis with a platform-specific master equation, invariant-based inverse
engineering of the frequency profile under bandwidth and positivity constraints,
geodesic and multi-tone control on the \(SU(1,1)\) manifold, and the extension to
larger and more general oscillator networks.

\section*{Acknowledgments}
H.H.H. acknowledges support from SECIHTI grant CBF-2023-2024-1937 and Sabbatical Grant 2025.\\

\section*{Author contributions and AI disclosure}
All authors contributed to the development, verification, and revision of this
work and approved the final manuscript. ChatGPT (OpenAI) was used to support
conceptual discussions, manuscript revision, and Python code development. All
AI-assisted material and results were critically reviewed and independently
verified by the authors, who take full responsibility for the work.

\section*{Data and code availability}

No external data are required.  The complete reproducibility package supplied
with this submission contains the Python source, environment requirements, and
scripts that generate every figure and quoted numerical value.  A public
persistent repository identifier will be added to the version of record.

\appendix

\section{Covariance moments and physical-basis transformation}
\label{app:cov}

For the invariant vacuum, Eq.~\eqref{eq:phi_n} with \(n=0\) is
\begin{equation}
	\phi_{0,k}(Q,t)=(\pi\hbar\rho_k^2)^{-1/4}
	\exp\!\left[-\frac{Q^2}{2\hbar\rho_k^2}
	+\frac{\ii\dot\rho_k}{2\hbar\rho_k}Q^2\right].
	\label{eq:app_ground}
\end{equation}
Writing the exponent as \(-\gamma Q^2\) with
\(\gamma=(1-\ii\rho_k\dot\rho_k)/(2\hbar\rho_k^2)\), Gaussian integration gives
\begin{equation}
	\langle Q_k^2\rangle=\frac{\hbar\rho_k^2}{2},\qquad
	\tfrac12\langle Q_kP_k+P_kQ_k\rangle=\frac{\hbar\rho_k\dot\rho_k}{2},\qquad
	\langle P_k^2\rangle=\frac{\hbar}{2}\bigl(\dot\rho_k^2+\rho_k^{-2}\bigr),
	\label{eq:app_moments}
\end{equation}
which is the block~\eqref{eq:modeCM}.  The cross moment is sometimes assigned a
spurious extra factor of two; the value in Eq.~\eqref{eq:app_moments} follows
directly from \(P=-\ii\hbar\,\partial_Q\) acting on
Eq.~\eqref{eq:app_ground}, and is the value required by
\(\det\CM_k=(\hbar/2)^2\).  The symplectic matrix induced by the fixed
rotation~\eqref{eq:normal}, in the ordering \((x_1,p_1,x_2,p_2)\), is
\begin{equation}
	S=\frac1{\sqrt2}\begin{pmatrix}
		1&0&1&0\\0&1&0&1\\1&0&-1&0\\0&1&0&-1
	\end{pmatrix},
	\label{eq:app_S}
\end{equation}
and \(\CM=S\,(\CM_+\oplus\CM_-)\,S^{\trans}\) yields
Eq.~\eqref{eq:physicalCM}.

\section{Derivation of the closed logarithmic negativity}
\label{app:negativity}

Set \(\hbar/2=1\) and let \(M_\pm\) be the dimensionless blocks
of~\eqref{eq:modeCM}, each with \(\det M_\pm=1\).  From
Eq.~\eqref{eq:physicalCM},
\begin{equation}
	\det A=\tfrac14(2+X),\qquad \det C=\tfrac14(2-X),
	\label{eq:app_dets}
\end{equation}
with \(X\) as in Eq.~\eqref{eq:X}; equivalently
\(X=\tr(M_+^{-1}M_-)\), using
\(\det(M_++M_-)=\det M_++\det M_-+\tr(M_+^{-1}M_-)=2+X\).  Partial transposition
multiplies the off-diagonal block on one side by \(\diag(1,-1)\), reversing the
sign of \(\det C\) in the two-mode invariant, so
\(\widetilde\Delta=\det A+\det B-2\det C=X\).  The global state is pure, hence
\(\det\CM=1\), and Eq.~\eqref{eq:nupt} becomes
\begin{equation}
	\left(\frac{2\widetilde\nu_-}{\hbar}\right)^2=\frac{X-\sqrt{X^2-4}}{2}.
	\label{eq:app_nu}
\end{equation}
Writing \(X/2=\cosh y\) (\(y\geq0\)) gives
\(X-\sqrt{X^2-4}=2e^{-y}\), so \((2\widetilde\nu_-/\hbar)^2=e^{-y}\), i.e.\
\(2\widetilde\nu_-/\hbar=e^{-y/2}\) and
\(\EN=-\ln(2\widetilde\nu_-/\hbar)=y/2\).  This proves
Eq.~\eqref{eq:closedEN} and fixes the exponent in
Eq.~\eqref{eq:DuanRelation}.  Equivalently, for a pure two-mode state
\(\EN=\operatorname{arcosh}\!\sqrt{\det A}=\operatorname{arcosh}\!\sqrt{(2+X)/4}
=\tfrac12\operatorname{arcosh}(X/2)\), which provides an independent check.

\section{Why the optimized Duan relation is restricted}
\label{app:duan}

Every pure exchange-symmetric two-mode Gaussian state is locally symplectically
equivalent to a two-mode squeezed vacuum with an effective parameter
\(r_{\rm eff}\), whose standard-form covariance matrix is
\begin{equation}
	A=B=\frac\hbar2\cosh(2r_{\rm eff})\,I_2,\qquad
	C=\frac\hbar2\sinh(2r_{\rm eff})\,\diag(1,-1).
	\label{eq:app_tmsv}
\end{equation}
Evaluating the EPR pair~\eqref{eq:EPR} on the squeezed quadratures gives
\begin{equation}
	\frac{\mathcal D_{\rm opt}}{\mathcal D_{\rm sep}}=e^{-2r_{\rm eff}},\qquad
	\frac{2\widetilde\nu_-}{\hbar}=e^{-2r_{\rm eff}},\qquad
	\EN=2r_{\rm eff},
	\label{eq:app_duanrel}
\end{equation}
which is Eq.~\eqref{eq:DuanRelation}.  The step that fails for mixed or
asymmetric states is the reduction to a single effective parameter: the local
normal form then carries additional symplectic invariants, and a single EPR
variance no longer determines \(\EN\).  Hence \(W_D\) is an optimized
inseparability witness for the family studied here, not a universal entanglement
measure.

\section{Algebraic Lewis--Riesenfeld phase}
\label{app:phase}

The eigenvalue equation for \(I_k\) follows from Eq.~\eqref{eq:Inumber}.  The
Lewis--Riesenfeld theorem states that, for a nondegenerate invariant basis, the
phase obeys
\begin{equation}
	\hbar\,\dot\alpha_n(t)=\langle n,t|\,\ii\hbar\partial_t-H(t)\,|n,t\rangle,
	\label{eq:app_LR}
\end{equation}
together with the off-diagonal condition implied by \(\dot I_k=0\),
\begin{equation}
	\langle m,t|\,\ii\hbar\partial_t-H(t)\,|n,t\rangle=0\qquad(m\neq n),
	\label{eq:app_offdiag}
\end{equation}
which is the correct statement --- not \(\langle m,t|\partial_t|n,t\rangle=0\)
separately.  Substituting the explicit states~\eqref{eq:phi_n} and using the
Ermakov equation~\eqref{eq:Ermakov} yields
\(\dot\alpha_{n,k}=-(n+\tfrac12)/\rho_k^2\), i.e.\ the phase of
Eq.~\eqref{eq:LRphase}.  This avoids the common error of confusing the ordinary
time derivative of a time-dependent invariant with its total Heisenberg
derivative.

\bibliographystyle{unsrtnat}
\bibliography{references}

@article{LewisRiesenfeld1969,
  author  = {Lewis, H. R. and Riesenfeld, W. B.},
  title   = {An Exact Quantum Theory of the Time-Dependent Harmonic Oscillator and of a Charged Particle in a Time-Dependent Electromagnetic Field},
  journal = {Journal of Mathematical Physics},
  volume  = {10},
  pages   = {1458--1473},
  year    = {1969},
  doi     = {10.1063/1.1664991}
}

@article{Pinney1950,
  author  = {Pinney, Edmund},
  title   = {The Nonlinear Differential Equation $y''+p(x)y+cy^{-3}=0$},
  journal = {Proceedings of the American Mathematical Society},
  volume  = {1},
  pages   = {681},
  year    = {1950},
  doi     = {10.1090/S0002-9939-1950-0037979-4}
}

@article{Duan2000,
  author  = {Duan, Lu-Ming and Giedke, G. and Cirac, J. I. and Zoller, P.},
  title   = {Inseparability Criterion for Continuous Variable Systems},
  journal = {Physical Review Letters},
  volume  = {84},
  pages   = {2722--2725},
  year    = {2000},
  doi     = {10.1103/PhysRevLett.84.2722}
}

@article{Simon2000,
  author  = {Simon, R.},
  title   = {Peres--Horodecki Separability Criterion for Continuous Variable Systems},
  journal = {Physical Review Letters},
  volume  = {84},
  pages   = {2726--2729},
  year    = {2000},
  doi     = {10.1103/PhysRevLett.84.2726}
}

@article{VidalWerner2002,
  author  = {Vidal, G. and Werner, R. F.},
  title   = {Computable Measure of Entanglement},
  journal = {Physical Review A},
  volume  = {65},
  pages   = {032314},
  year    = {2002},
  doi     = {10.1103/PhysRevA.65.032314}
}

@article{Weedbrook2012,
  author  = {Weedbrook, Christian and Pirandola, Stefano and Garc{\'i}a-Patr{\'o}n, Ra{\'u}l and Cerf, Nicolas J. and Ralph, Timothy C. and Shapiro, Jeffrey H. and Lloyd, Seth},
  title   = {Gaussian Quantum Information},
  journal = {Reviews of Modern Physics},
  volume  = {84},
  pages   = {621--669},
  year    = {2012},
  doi     = {10.1103/RevModPhys.84.621}
}

@article{Urzua2019,
  author  = {Urz{\'u}a, Alejandro R. and Ramos-Prieto, Ir{\'a}n and Fern{\'a}ndez-Guasti, Manuel and Moya-Cessa, H{\'e}ctor M.},
  title   = {Solution to the Time-Dependent Coupled Harmonic Oscillators Hamiltonian with Arbitrary Interactions},
  journal = {Quantum Reports},
  volume  = {1},
  pages   = {82--90},
  year    = {2019},
  doi     = {10.3390/quantum1010009}
}

@article{Park2019,
  author  = {Park, DaeKil},
  title   = {Dynamics of Entanglement in Three Coupled Harmonic Oscillator System with Arbitrary Time-Dependent Frequency and Coupling Constants},
  journal = {Quantum Information Processing},
  volume  = {18},
  pages   = {282},
  year    = {2019},
  doi     = {10.1007/s11128-019-2393-4}
}

@article{Abidi2021,
  author  = {Abidi, A. and Trabelsi, A. and Krichene, S.},
  title   = {Coupled Harmonic Oscillators and Their Application in the Dynamics of Entanglement and the Nonadiabatic Berry Phases},
  journal = {Canadian Journal of Physics},
  volume  = {99},
  number  = {10},
  pages   = {898--906},
  year    = {2021},
  doi     = {10.1139/cjp-2020-0410}
}

@article{Tobalina2020,
  title={Invariant-based inverse engineering of time-dependent, coupled harmonic oscillators},
  author={Tobalina, Ander and Torrontegui, E and Lizuain, Ion and Palmero, Mikel and Muga, Juan Gonzalo},
  journal={Physical Review A},
  volume={102},
  number={6},
  pages={063112},
  year={2020},
  publisher={APS},
  doi={10.1103/PhysRevA.102.063112}
}

@misc{Ghaba2026,
  author        = {Ghaba, Ayoub and Hab-arrih, Radouan and Atmani, Elhoussine and Slaoui, Abdallah},
  title         = {Dynamics of Quantum Entanglement in Two Time-Dependent Coupled Harmonic Oscillators},
  year          = {2026},
  eprint        = {2606.29617},
  archivePrefix = {arXiv},
  primaryClass  = {quant-ph},
  doi           = {10.48550/arXiv.2606.29617}
}

@article{Leibfried2003,
  author  = {Leibfried, D. and Blatt, R. and Monroe, C. and Wineland, D.},
  title   = {Quantum Dynamics of Single Trapped Ions},
  journal = {Reviews of Modern Physics},
  volume  = {75},
  pages   = {281--324},
  year    = {2003},
  doi     = {10.1103/RevModPhys.75.281}
}

@article{Lo2015,
  author  = {Lo, Hsiang-Yu and Kienzler, Daniel and de Clercq, Ludwig and Marinelli, Matteo and Negnevitsky, Vlad and Keitch, Ben C. and Home, Jonathan P.},
  title   = {Spin--Motion Entanglement and State Diagnosis with Squeezed Oscillator Wavepackets},
  journal = {Nature},
  volume  = {521},
  pages   = {336--339},
  year    = {2015},
  doi     = {10.1038/nature14458}
}

@article{Maslennikov2019,
  author  = {Maslennikov, Gleb and Ding, Shiqian and Habl{\"u}tzel, Roland and Gan, Jaren and Roulet, Alexandre and Nimmrichter, Stefan and Dai, Jibo and Scarani, Valerio and Matsukevich, Dzmitry},
  title   = {Quantum Absorption Refrigerator with Trapped Ions},
  journal = {Nature Communications},
  volume  = {10},
  pages   = {202},
  year    = {2019},
  doi     = {10.1038/s41467-018-08090-0}
}

@article{Menon2026,
  author  = {Menon, Abhishek and Straus, Michael and Tomaras, George and Jeanette, Liam and Sheffield, April X. and Valdez, Devon and Xie, Yuanheng and So, Visal and Luo, De and Suganthi, Midhuna Duraisamy and Dugan, Mark and Bado, Philippe and Linke, Norbert M. and Pagano, Guido and Zhuravel, Roman},
  title   = {Monolithic Segmented Three-Dimensional Ion Trap for Quantum Technology Applications},
  journal = {Physical Review Applied},
  volume  = {26},
  pages   = {044037},
  year    = {2026},
  doi     = {10.1103/srq8-cgnt}
}

@article{Aspelmeyer2014,
  author  = {Aspelmeyer, Markus and Kippenberg, Tobias J. and Marquardt, Florian},
  title   = {Cavity Optomechanics},
  journal = {Reviews of Modern Physics},
  volume  = {86},
  pages   = {1391--1452},
  year    = {2014},
  doi     = {10.1103/RevModPhys.86.1391}
}

@article{Devoret2013,
  author  = {Devoret, M. H. and Schoelkopf, R. J.},
  title   = {Superconducting Circuits for Quantum Information: An Outlook},
  journal = {Science},
  volume  = {339},
  pages   = {1169--1174},
  year    = {2013},
  doi     = {10.1126/science.1231930}
}

@article{Mirkhalaf2026,
  title={Frequency shifts heralding ground-state squeezing and entanglement of two coupled harmonic oscillators},
  author={Mirkhalaf, Safoura and Ritsch, Helmut and Gietka, Karol},
  journal={Physical Review A},
  volume={114},
  number={2},
  pages={022426},
  year={2026},
  publisher={APS},
  doi={10.1103/sfy4-nx52}
}

@article{GonzalezHenao2015,
  author  = {Gonzalez-Henao, J. C. and Pugliese, E. and Euzzor, S. and Abdalah, S. F. and Meucci, R. and Roversi, J. A.},
  title   = {Generation of Entanglement in Quantum Parametric Oscillators Using Phase Control},
  journal = {Scientific Reports},
  volume  = {5},
  pages   = {13152},
  year    = {2015},
  doi     = {10.1038/srep13152}
}

@article{Buchmann2018,
  author  = {Buchmann, L. F. and M{\o}lmer, K. and Petrosyan, D.},
  title   = {Controllability in Tunable Chains of Coupled Harmonic Oscillators},
  journal = {Physical Review A},
  volume  = {97},
  pages   = {042111},
  year    = {2018},
  doi     = {10.1103/PhysRevA.97.042111}
}

@article{Plenio2005,
 author = {Plenio, M. B.},
  title = {Logarithmic Negativity: A Full Entanglement Monotone That is not Convex},
  journal = {Phys. Rev. Lett.},
  volume = {95},
  issue = {9},
  pages = {090503},
  numpages = {4},
  year = {2005},
  month = {Aug},
  publisher = {American Physical Society},
  doi = {10.1103/PhysRevLett.95.090503}
}

\end{document}